\documentstyle[11pt,epsfig,amsbsy]{article}

\def\be{\begin{equation}}
\def\ee{\end{equation}}
\def\ba{\begin{eqnarray}}
\def\ea{\end{eqnarray}}

\def\ga{\mathrel{\raise.3ex\hbox{$>$\kern-.75em\lower1ex\hbox{$\sim$}}}}
\def\la{\mathrel{\raise.3ex\hbox{$<$\kern-.75em\lower1ex\hbox{$\sim$}}}}

\newcommand{\fr}[2]{\frac{#1}{#2}}

\newcommand{\m}{\rm{m}}

\newcommand{\omde}{\omega_{\rm{de}}}
\newcommand{\Omo}{\Omega_{\rm{m}}^{0}}
\newcommand{\Odeo}{\Omega_{\rm{de}}^{0}}

\newcommand{\rde}{\rho_{\rm{de}}}
\newcommand{\rhom}{\rho_{\rm{m}}}
\newcommand{\rcr}{\rho_{\rm{cr}}}

\newcommand{\ws}{\rm{ws}}

\newcommand{\vir}{\rm{vir}}
\newcommand{\ta}{\rm{ta}}
\newcommand{\cta}{\rm{cta}}
\newcommand{\de}{\rm{de}}
\newcommand{\mc}{\rm{cluster}}
\newcommand{\rhoc}{\rho_{\rm{cluster}}}
\newcommand{\dec}{\rm{dec}}

\newcommand{\AS}{\rm{ArcSin}}
\newcommand{\EF}{\rm{EllipticF}}
\newcommand{\EPi}{\rm{EllipticPi}}
\newcommand{\sk}{\rm{sk}}

\newcommand{\analy}{\rm{analy}}
\newcommand{\EdS}{\rm{EdS}}
\begin{document}

\baselineskip=16pt
\begin{titlepage}
\begin{center}

\vspace{0.5cm}

\large {\bf Spherical collapse model with non-clustering dark energy}
\vspace*{5mm} \normalsize

{\bf Seokcheon Lee$^{\,1,2}$ and Kin-Wang Ng$^{\,1,2,3}$}

\smallskip
\medskip

$^1${\it Institute of Physics, Academia Sinica, \\
Taipei, Taiwan 11529, R.O.C.}

$^2${\it Leung Center for Cosmology and Particle Astrophysics, National Taiwan University, \\ Taipei, Taiwan 10617, R.O.C.}

$^3${\it Institute of Astronomy and Astrophysics, \\
 Academia Sinica, Taipei, Taiwan 11529, R.O.C.}

\smallskip
\end{center}

\vskip0.6in

\centerline{\large\bf Abstract}

We investigate a spherical overdensity model for the non-clustering dark energy (DE) with the constant equation of state, $\omde$ in a flat universe. In this case, the exact solution for the evolution of the scale factor is obtained for general $\omde$. We also obtain the exact (when $\omde = -\fr{1}{3}$) and the approximate (when $\omde \neq -\fr{1}{3}$) solutions for the ratio of the overdensity radius to its value at the turnaround epoch ($y \equiv \fr{R}{R_{\ta}}$) for general cosmological parameters. Also the exact and approximate solutions of the overdensity at the turnaround epoch $\zeta = \fr{\rho_{\mc}}{\rho_{\m}} \Bigl|_{z=z_{\ta}}$ are obtained for general $\omde$. Thus, we are able to obtain the non-linear overdensity $\Delta \equiv 1 + \delta_{\rm{NL}} = \zeta \Bigl(\fr{x}{y} \Bigr)^3$ at any epoch for the given DE model. The non-linear overdensity at the virial epoch $\Delta_{\vir} = \zeta \Bigl(\fr{x_{\vir}}{y_{\vir}} \Bigr)^3$ is obtained by using the virial theorem and the energy conservation. The non-linear overdensity of every DE model converges to that of the Einstein de Sitter universe $\Delta_{\vir}^{\rm{EdS}} = 18 \pi^2 \Bigr(\fr{1}{2\pi} + \fr{3}{4} \Bigr)^2 \simeq 147$ when $z_{\vir}$ increases. We find that the observed quantities at high redshifts are insensitive to the different $\omde$ models. The low-redshift cluster ($z_{\vir} \sim 0.04$, {\it i.e.}, $z_{\ta} \sim 0.7$) shows the most model dependent feature and it should be a suitable object for testing DE models. Also as $\Omo$ increases, the model dependence of the observed quantities decreases. The error in the approximate solutions is at most 2 \% for a wide range of the parameter space. Even though the analytic forms of $y$ and $\zeta$ are obtained for the constant $\omde$, they can be generalized to the slowly varying $\omde$. Thus, these analytic forms of the scale factor, $y$, and $\zeta$ provide a very accurate and useful tool for measuring the properties of DE.

\vspace*{2mm}

\end{titlepage}

\section{Introduction}
\setcounter{equation}{0}

Galaxy clusters are the largest virialized structure formed by the gravitational collapse of small density perturbations. The spherical collapse model (SCM) is a simple and powerful tool for investigating the evolution of nonlinear structure and bound systems in the Universe \cite{Gunn,Peebles}.

Since the formation of the large scale dark matter potential well of clusters depends only on gravitation, the property of dark energy (DE) and the cosmological parameters can be constrained from the growth of large scale structure and the abundances of rich clusters of galaxies. There have been numerous papers investigating this problem with the spherical collapse model \cite{0401504,9604141,WS,0409481,0504465,0505308,0507257,07080868,09063349,09081333}.

In order to prohibit the spherical overdensity from collapsing to a singular point in a finite time, we need to use the virialization in the SCM. The common mistake in literatures is that one uses the correct value of the finite virial radius of the overdensity from the virial theorem but still uses the incorrect value of the background scale factor for an Einstein de Sitter (EdS) universe. However, this leads to an incorrect conclusion that the bound systems are formed when the nonlinear overdensity is about $178$. Equally, this leads to the incorrect critical linear threshold density $1.68$ \cite{sky}. When we use the correct values of the virial radius and the scale factor at the virial epoch, these values are lowered to $147$ and $1.58$, respectively.

In the next section, we review and analyze the SCM model in a flat universe including the non-clustering DE with the constant equation of state $\omde$. We obtain the exact solution for the evolution of the scale factor (a) in terms of the ratio of it to its value at the turnaround epoch ($x \equiv \fr{a}{a_{\ta}}$). We also obtain the exact (when $\omde = -\fr{1}{3}$) and the approximate (when $\omde \neq -\fr{1}{3}$) solutions for the ratio of the overdensity radius to its value at the turnaround epoch ($y \equiv \fr{R}{R_{\ta}}$) for general cosmological parameters. Also the exact and approximate solutions of the overdensity at the turnaround epoch $\zeta = \fr{\rho_{\mc}}{\rho_{\m}} \Bigl|_{z=z_{\ta}}$ are obtained for general $\omde$. The given approximate solutions have the less than $1.5$ \% errors over the parameter space in the concordance model. In Sec. $3$, we use the virial theorem to obtain the correct values of $x$ and $y$ at the virial epoch. We compare the values of the virial epoch and the nonlinear overdensity for the different models. We conclude in Sec. $4$.

\section{Spherical Collapse Model}
\setcounter{equation}{0}

Background evolution equations of the physical quantities in a flat FRW universe with the matter and the DE are given by
\ba && H^2 = \Bigl( \fr{\dot{a}}{a} \Bigr)^2 = \fr{8\pi G}{3}(\rhom + \rde) = \fr{8 \pi G}{3} \rcr \, , \label{H} \\ && \fr{\ddot{a}}{a} = - \fr{4 \pi G}{3} \Bigl[ \rhom + (1+ 3\omde) \rde \Bigr] \, , \label{ddota} \\ && \dot{\rho}_{\rm{m}} + 3 \Bigl( \fr{\dot{a}}{a} \Bigr) \rhom = 0 \, , \label{rhomdot} \\ && \dot{\rho}_{\rm{de}} + 3 ( 1 + \omde) \Bigl( \fr{\dot{a}}{a} \Bigr) \rde = 0 \, , \label{rdedot}\ea where $a$ is the scale factor, $\omde$ is the equation of state (eos) of the dark energy, $\rcr$ is the critical energy density, and $\rhom$ and $\rde$ are the energy densities of the matter and the dark energy, respectively. We consider the constant $\omde$ models only.

We investigate a spherical perturbation in the matter density \cite{Gunn,Peebles}. $\rhoc$ is the matter density within the spherical overdensity radius $R$. The flatness condition is not held because of the perturbation in the matter. Thus, we have another set of equations governing the dynamics of the spherical perturbation \cite{0401504} :
\ba && \fr{\ddot{R}}{R} = - \fr{4 \pi G}{3} \Bigl[ \rhoc + (1+ 3\omde) \rho_{\rm{dec}} \Bigr]  \, , \label{ddotR} \\ && \dot{\rho}_{\mc} + 3 \Bigl( \fr{\dot{R}}{R} \Bigr) \rhoc = 0 \, , \label{rhocdot} \\ && \dot{\rho}_{\rm{dec}} + 3 ( 1 + \omde) \Bigl( \fr{\dot{R}}{R} \Bigr) \rho_{\rm{dec}} = \alpha \Gamma \, , \nonumber \\
&& {\rm where} \,\, \Gamma = 3 ( 1 + \omde) \Bigl( \fr{\dot{R}}{R} - \fr{\dot{a}}{a} \Bigr) \rho_{\rm{dec}} \hspace{0.2in} \rm{with} \,\, 0 \leq \alpha \leq 1 \, , \label{rhodecdot} \ea where $\rho_{\rm{dec}}$ is the DE density within the cluster. $\alpha = 0$ gives the clustering behavior of DE and $\alpha = 1$ means the homogenous DE satisfying Eq. (\ref{rdedot}). For $\alpha = 1$ case, the system does not conserve energy as it collapses from the turnaround to the virialized state. We can solve Eq. (\ref{rhodecdot}) for the constant $\alpha$, \be \rho_{\dec} (a, R) = \rho_{\dec}^{\ta} \Biggl( \fr{a}{a_{\ta}} \Biggr)^{-3(1+\omde) \alpha} \Biggl( \fr{R}{R_{\ta}} \Biggr)^{-3(1+\omde)(1 - \alpha)} \, , \label{rhodec} \ee where $a_{\ta}$ and $R_{\ta}$ define the scale factor and the radius at the turnaround redshift $z_{\ta}$. There is an ambiguity of $\rho_{\dec}$ in Eq. (\ref{rhodec}) because it is a function of both $a$ and $R$ except when $\alpha = 1$ or $0$. Independent of $\alpha$, the radius of the overdensity $R$ evolves slower than the scale factor $a$ and reaches its maximum size $R_{\ta}$ at $z_{\ta}$ and then the system begins to collapse.

Now we adopt the notations in Ref. \cite{WS} to investigate the evolutions of $a$ and $R$, \ba x &=& \fr{a}{a_{\ta}} \, , \label{x} \\ y &=& \fr{R}{R_{\ta}} \, . \label{y} \ea Then from Eqs. (\ref{H}) and (\ref{ddotR}), we obtain \ba \fr{dx}{d \tau} &=& \fr{1}{\sqrt{x \Omega_{\rm{m}}(x)}} = \sqrt{x^{-1} + \fr{1}{Q_{\ta}} x^{-3 \omde -1}} \, , \label{dx} \\ \fr{d^2 y}{d \tau^2} &=& -\fr{1}{2} \Biggl[ \zeta y^{-2} + (1 + 3 \omde) \fr{1}{Q_{\cta}} x^{-3(1+\omde) \alpha} y^{-3(1+\omde)(1-\alpha) +1} \Biggr] \, , \label{ddy} \ea
where $d \tau \equiv H_{\ta} \sqrt{\Omega_{\rm{m}}(x_{\ta})} dt$, $\zeta \equiv \fr{\rho_{\mc}}{\rho_{\m}} |_{z_{\ta}}$, $Q_{\ta} \equiv \fr{\rho_{\m}}{\rho_{\de}}|_{z_{\ta}} = \fr{\Omo}{\Odeo} (1+z_{\ta})^{-3\omde}$, $Q_{\cta}\equiv \fr{\rho_{\m}}{\rho_{\dec}}|_{z_{\ta}} = \fr{\Omo}{\Omega_{\dec}^{0}} (1+z_{\ta})^{3 - 3 (1 + \omde) \alpha } \,
\Bigl( \fr{R_{0}}{R_{\ta}} \Bigr)^{-3(1+\omde)(1-\alpha)}$, and $x_{\ta} = 1$ from Eq. (\ref{x}). $\Omo$ and $\Odeo$ represent the present values of the energy density contrasts of the matter and the DE, respectively. $Q_{\cta}$ becomes $Q_{\ta}$ when $\alpha = 1$ (non-clustering DE). 
Also when $\omde = -1/3$ or $-1$, Eq. (\ref{ddy}) becomes independent of $\alpha$. We can solve Eq. (\ref{ddy}) analytically in these cases. We limit our consideration to only the non-clustering DE.

We are able to obtain the analytic solution of Eq. (\ref{dx}) \be \int_{0}^{x} dx' \sqrt{x' \Omega_{\rm{m}}(x')} = \int_{0}^{\tau} d \tau^{'} \,\, \Rightarrow \,\, \fr{2}{3} x^{\fr{3}{2}} F \Bigl[ \fr{1}{2}, -\fr{1}{2 \omde}, 1 - \fr{1}{2 \omde}, - \fr{x^{-3 \omde}}{Q_{\ta}} \Bigr] = \tau \, , \label{xtau} \ee where $F$ is the hypergeometric function and we use the boundary condition $x = 0$ when $\tau = 0$ (see Appendix for details). This solution is the same as the one in Ref. \cite{sky} when $\omde = -\fr{1}{3}$. The above equation gives the well known relation $a(t) \propto \rm{sinh}^{\fr{2}{3}} \Bigl[\fr{3}{2} \sqrt{\Odeo} H_0 t \Bigr]$ when $\omde = -1$.  From this equation we are able to find the exact turnaround time $\tau_{\ta}$ for general $\omde$ models, \be \tau_{\ta} = \fr{2}{3} F \Bigl[  \fr{1}{2}, -\fr{1}{2 \omde}, 1 - \fr{1}{2 \omde}, - Q_{\ta}^{-1} \Bigr] \, , \label{tautaw} \ee where we use the fact that $x_{\ta} = 1$. It is useful to have the exact analytic form of the turnaround time to investigate the accuracies of both numerical and analytical calculations.

\begin{center}
\begin{figure}
\vspace{1.5cm}
\centerline{
\psfig{file=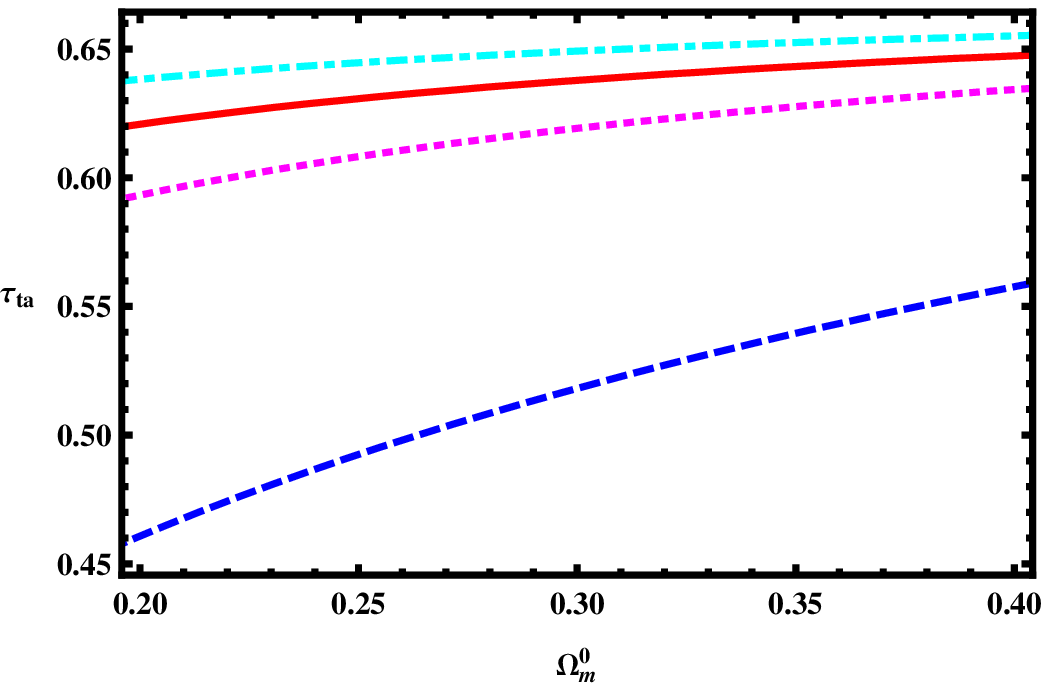, width=6.5cm} \psfig{file=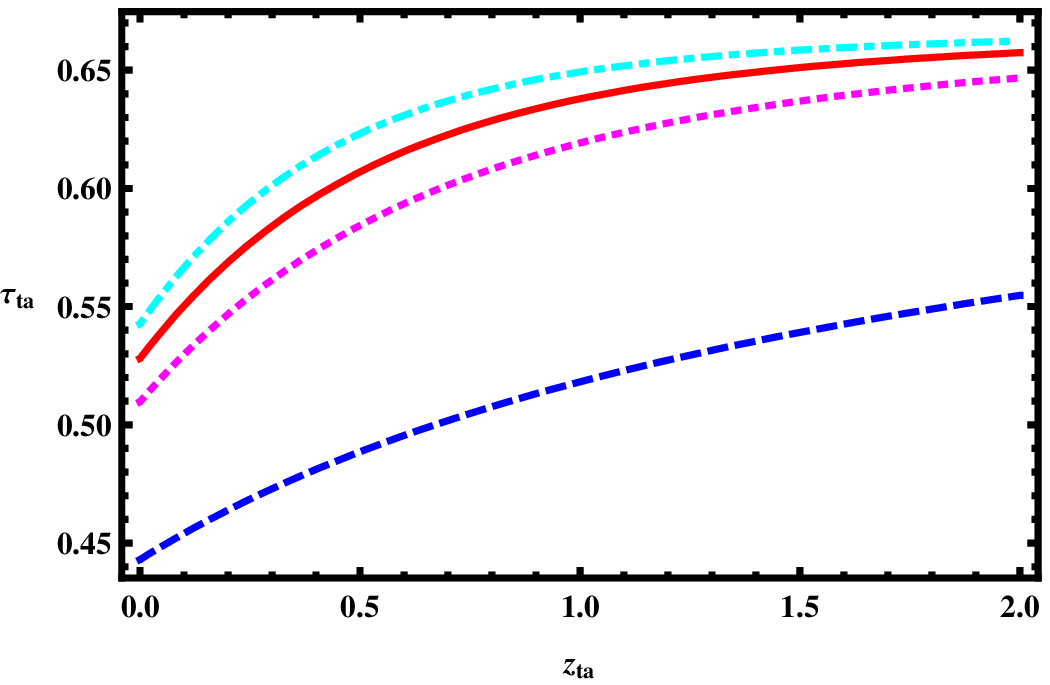, width=6.5cm} }
\vspace{-0.5cm}
\caption{ $\tau_{\ta}$ for the different values of $\omde$. a) $\tau_{\ta}$ v.s. $\Omo$ for the different values of $\omde = -1.2, -1.0, -0.8$, and $-1/3$ (from top to bottom) when $z_{\ta} = 1.0$. b) $\tau_{\ta}$ v.s. $z_{\ta}$ for the same models when we choose $\Omo = 0.3$.} \label{fig1}
\end{figure}
\end{center}

As expected, $\tau_{\ta}$ depends on $\omde$, $\Omo$, and $z_{\ta}$ as given in Eq. (\ref{tautaw}). We show these properties of $\tau_{\ta}$ in Fig. \ref{fig1}. In the left panel of Fig. \ref{fig1}, we show $\tau_{\ta}$ as a dependence on the values of $\Omo$ for the different DE models. We choose $z_{\ta} = 1$ in this figure. The dot-dashed, solid, dotted, and dashed lines (from top to bottom) correspond to $\omde = -1.2, -1.0, -0.8$, and $-1/3$, respectively. If we have the same $\Omo$ for the different models, then it takes longer time to the turnaround for the smaller values of $\omde$. The smaller value of $\omde$ means the higher pressure against the gravity. Thus, it takes longer time to reach to the maximum radius of the overdensity for the higher pressure universe when we have the same amount of $\Omo$. We also need more matter to overcome the higher pressure. We also show the $z_{\ta}$ dependence of $\tau_{\ta}$ for the different DE models when we choose $\Omo = 0.3$ in the right panel of Fig. \ref{fig1}. Again the smaller values of $\omde$ give the larger $\tau_{\ta}$. An interesting feature in this figure is that as $z_{\ta}$ increases so does $Q_{\ta}$. For large $z_{\ta}$, $Q_{\ta}^{-1}$ approaches to $0$ and thus $\tau_{\ta}$ approaches to $\fr{2}{3} F \Bigl[ \fr{1}{2}, -\fr{1}{2 \omde}, 1 - \fr{1}{2 \omde}, 0 \Bigr] = \fr{2}{3} \sim 0.67$ independent of $\omde$.

Now we work on the evolution of $y$. We only consider the $\alpha = 1$ case and Eq. (\ref{ddy}) becomes \be \fr{d^2 y}{d \tau^2} = -\fr{1}{2} \Biggl[ \zeta y^{-2} + (1 + 3 \omde) \fr{1}{Q_{\ta}} x^{-3(1+\omde)} y \Biggr] \, .  \label{ddy2} \ee For general $\omde$, there is no analytic solution of Eq. (\ref{ddy2}). However, we are able to obtain the exact analytic solutions for $\omde = -1/3$ and $-1$. First, we consider the case when $\omde = -\fr{1}{3}$. In this case, the second term in the right hand side of Eq. (\ref{ddy2}) disappears and the equation becomes \cite{HB}
\be \fr{d y^2}{d \tau^2} =-\fr{1}{2} \zeta y^{-2} \equiv f(y) \,\, \Rightarrow \,\, \int_{0}^{y} \fr{dy'}{\sqrt{-c_{1} + 2 \int^{y'} f(y'') dy''}} = c_{2} \pm \tau \, . \label{y13} \ee The solutions of the above equation are given by (see Ref. \cite{sky}) \ba && \AS [\sqrt{y} ] - \sqrt{y(1-y)} = \sqrt{\zeta_{13}} \tau \,\, , {\rm when} \,\, \tau \leq \tau_{\ta 13} \, ,  \label{y133l} \\ && \sqrt{y(1-y)} - \AS [\sqrt{y} ] + \fr{\pi}{2} = \sqrt{\zeta_{13}} ( \tau - \tau_{\ta 13}) \,\, , {\rm when} \tau \geq \tau_{\ta 13} \, , \label{y133}  \ea where $\zeta_{13}$ and $\tau_{\ta 13}$ denote respectively $\zeta$ and $\tau_{\ta}$ for $\omde = -\fr{1}{3}$. Even though the above solutions given in Eqs. (\ref{y133l}) and (\ref{y133}) are only true for $\omde = -\fr{1}{3}$, these solutions can be good approximate solutions for general $\omde$ when we consider the high-redshift clusters. It is because the high redshift clusters which are formed at the high $z_{\vir}$ require the high $z_{\ta}$ in each model. As $z_{\ta}$ increases, so does $Q_{\ta}$ for the fixed value of $\Omo \sim 0.3$. Thus, the second term in Eq. (\ref{ddy2}) is negligible compared to the first one and this equation approximately equals to Eq. (\ref{y13}). Also these solutions are the same as the ones in the Einstein de Sitter (EdS) universe ($\Omo = 1$ and $\Odeo = 0$). It is because in the EdS universe, $Q_{\ta}^{-1} = 0$ and the evolution equation of $y$ is the same as in Eq. (\ref{y13}). Also the solution given in Eq. (\ref{y133l}) is a good approximate solution for general $\omde$ when we consider the evolution of $y$ at early epoch. As $y \rightarrow 0$, the second term in Eq. (\ref{y13}) is negligible again.
\begin{center}
\begin{figure}
\vspace{1.5cm}
\centerline{
\psfig{file=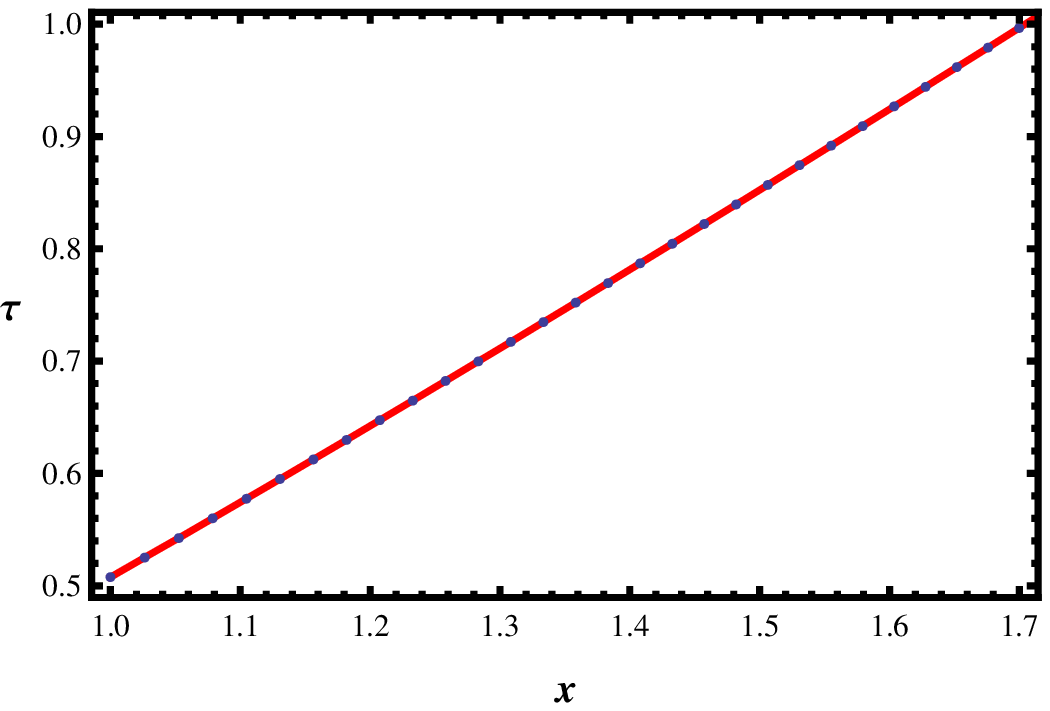, width=6.5cm} \psfig{file=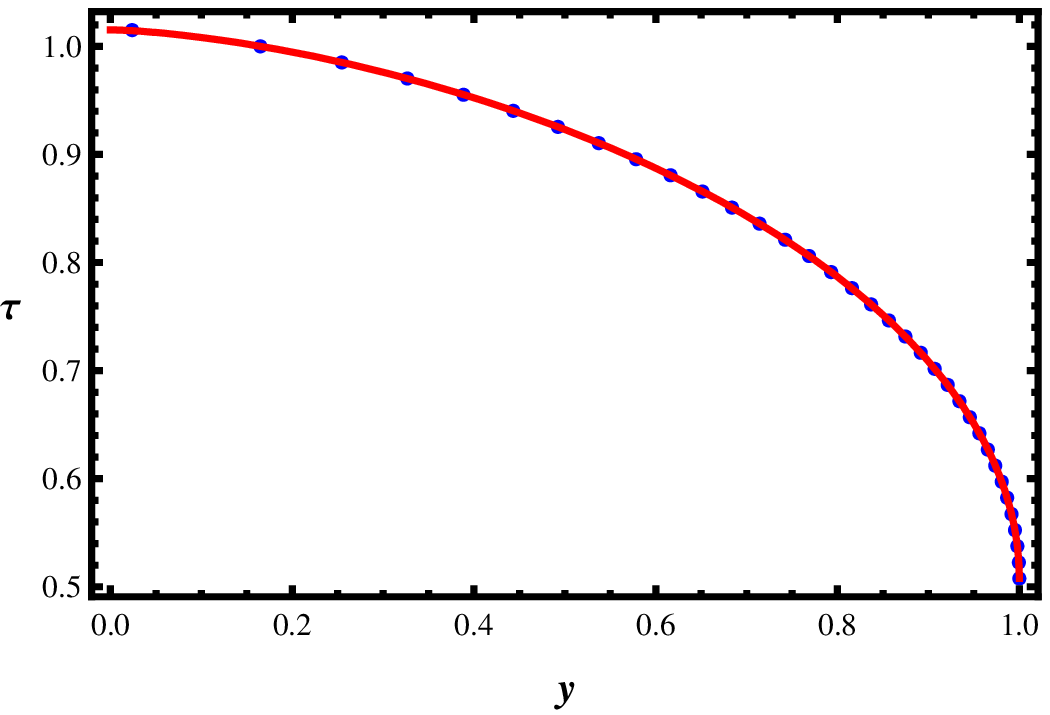, width=6.5cm} }
\vspace{-0.5cm}
\caption{ Both numerical and analytic solutions, denoted by dots and solid lines
    respectively. a) Cosmological evolutions of $\tau$ for $\omde = -\fr{1}{3}$ as a function of $x$ when we choose $z_{\ta} = 0.8$ and $\Omo = 0.3$. b) Evolutions of $\tau$ as a function of $y$ for the same values of $z_{\ta}$ and $\Omo$ as in a).} \label{fig2}
\end{figure}
\end{center}

As long as we know the exact virialized epoch $z_{\vir}$, Eq. (\ref{y133}) provides the exact value of $\zeta$ for the different values of $\Omo$ and $z_{\ta}$ when $\omde = -\fr{1}{3}$. We will investigate $z_{\vir}$ based on the virial theorem later. However, we first consider the simpler estimation to obtain $\zeta$ values for the general $\omde$. If we define $z_{\rm{c}}$ as the redshift at which the cluster formally collapses to $R = 0$ according to a spherical solution, then this collapse time is twice the turnaround time $t_{\rm{c}} = 2 t_{\ta}$ ({\it i.e.} $\tau(z_{\rm{c}}) = 2 \tau(z_{\ta})$) \cite{WS}. If we adopt this fact, then we are able to obtain $\zeta$ from the above analytic solution in Eq. (\ref{y133}) \be \zeta_{13} = \Biggl( \fr{\pi/2}{\fr{2}{3} F \Bigl[ \fr{1}{2}, \fr{3}{2}, \fr{5}{2}, \fr{-1}{Q_{\ta 13}} \Bigr]} \Biggr)^2 = \Biggl( \fr{3 \pi}{4} \Biggr)^2 \Biggl(F \Bigl[ \fr{1}{2}, \fr{3}{2}, \fr{5}{2}, -\fr{(1 - \Omo)}{\Omo} (1 + z_{\ta}) ^{-1} \Bigr] \Biggr)^{-2} \, . \label{zeta} \ee When $\Omo = 1$, $F \Bigl[ \fr{1}{2}, \fr{3}{2}, \fr{5}{2}, 0 \Bigr] = 1$. Thus, $\zeta_{13} = (\fr{3 \pi}{4})^2$ when $\Omo = 1$. This factor $(\fr{3 \pi}{4})^2$ is the well-known value of $\zeta$ for the EdS universe \cite{Peebles,Kihara}. The general value of $\zeta$ for $\omde = -\fr{1}{3}$ when $\Omo \neq 1$ is given by Eq. (\ref{zeta}). After we obtain the value of $\zeta_{13}$ given in Eq. (\ref{zeta}), we are able to find the values of $x$ and $y$ at any $\tau_{\ta} \leq \tau \leq \tau_{c}$ from Eqs. (\ref{xtau}) and (\ref{y133}). We constrain $\tau \geq \tau_{\ta}$ to show the evolution of $x$ after the overdensity radius reaches its maximum. In the left panel of Fig.~ \ref{fig2}, we show the evolution of $\tau$ as a function of $x$ when $z_{\ta} = 0.8$ and $\Omo = 0.3$. The numerical solution (denoted by dots) is definitely the same as the analytic solution (solid line). In the right panel of Fig. \ref{fig2}, we show the evolutions of both analytic and numerical solutions of $y(\tau)$ for the same values of $z_{\ta}$ and $\Omo$ as in the left panel of Fig. \ref{fig2}.

Now we consider Eq. (\ref{ddy2}) when $\omde = -1$. The differential equation of $y$ for $\omde = -1$ becomes \be \fr{d^2 y}{d \tau^2} = -\fr{1}{2} \Biggl[ \zeta y^{-2} - \fr{2}{Q_{\ta}} y \Biggr] \,\, \Rightarrow \,\,  \int^{y} \fr{dy'}{\sqrt{-c_{1} + \fr{\zeta}{y'} + \fr{y'^{2}}{Q_{\ta}}}} = c_{2} \pm \tau \, . \label{ddywL} \ee We obtain $c_{1} = \zeta + Q_{\ta}^{-1}$ from Eq. (\ref{ddywL}) by using the initial condition $\fr{dy}{d \tau}|_{z_{\ta}} = 0$. The solution of the above equation is given by (see Appendix for details) \be c_{F} \Biggl( \EF \Bigl[\AS [B], C \Bigr] + c_{P} \EPi \Bigl[ A, \AS [B], C \Bigr] \Biggr) = c_{2} - \tau \,\,\ , {\rm when} \,\, \tau \geq \tau_{\ta} \, , \label{ytauwL} \ee where EllipticF and EllipticPi correspond to the elliptic integral of the first kind and the incompelete elliptic integral of the third kind, respectively \cite{Abramowitz}. Also $c_{F} = - \fr{2 \sqrt{2 Q_{\ta}}}{\sqrt{2 Q_{\ta} \zeta - 1 + \sqrt{1 + 4 Q_{\ta} \zeta}}}$ and $c_{P} = - \fr{2 (Q_{\ta} \zeta - 2)}{(3 + \sqrt{1 + 4 Q_{\ta} \zeta}) }$. The integral constant $c_{2}$ is able to be obtained when we use the boundary condition $y(\tau_{ta}) = 1$. Even though this is the exact analytic solution of $y$ for $\omde = -1$, this formula has some problems. First, this is inconvenient to use. Second, this formula does not produce the correct value when $B > 1$. However, it is useful to have this analytic solution when we derive the form of fitting formulae for $\zeta$ and $y$ for general $\omde$ cases.
\begin{center}
\begin{figure}
\vspace{1.5cm}
\centerline{
\psfig{file=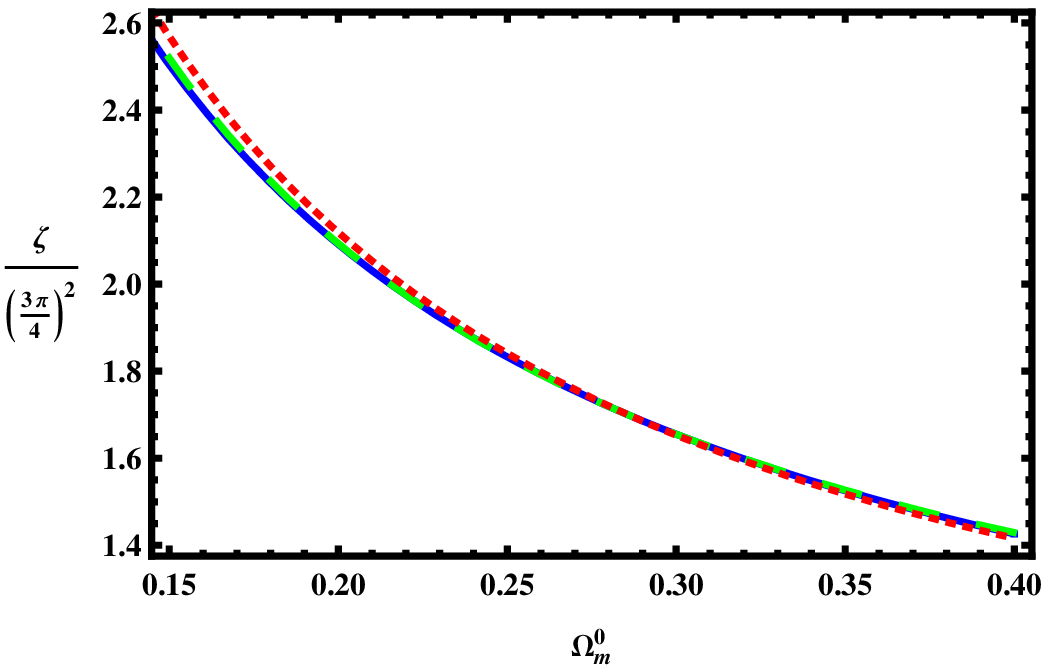, width=6.5cm} \psfig{file=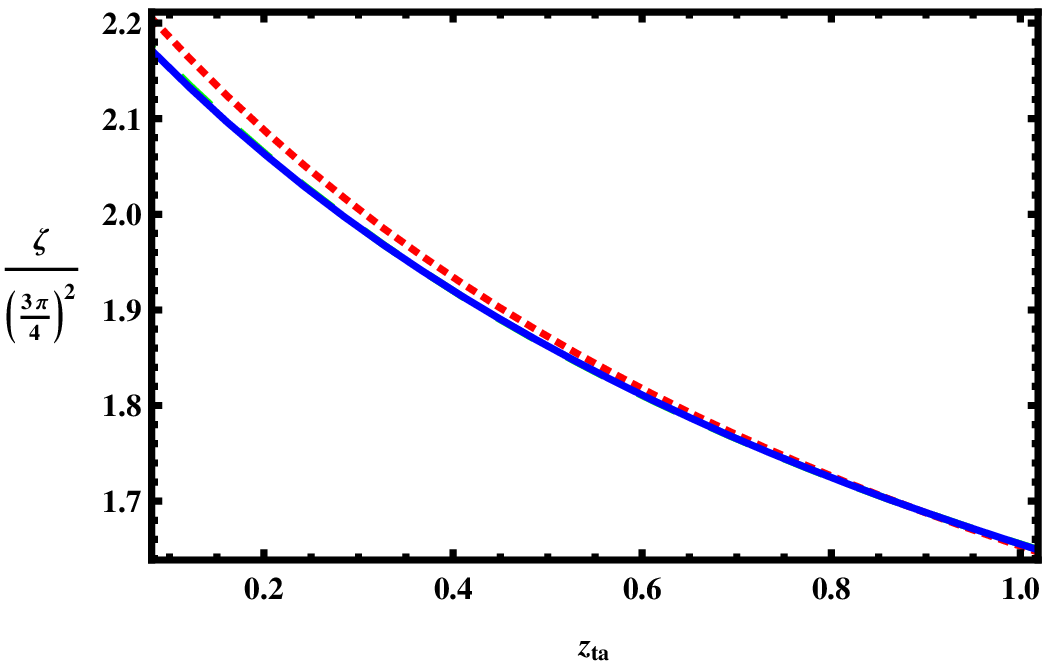, width=6.5cm} }
\vspace{-0.5cm}
\caption{ a) $\zeta_{\ws}, \zeta_{13}$, and $\zeta_{\sk}$ (from top to bottom) versus $\Omo$ when we choose $z_{\ta} = 1$. b) $\zeta$s versus $z_{\ta}$ when we choose $\Omo = 0.3$.} \label{fig3}
\end{figure}
\end{center}

$\zeta_{13}$ given in Eq. (\ref{zeta}) is exact only for $\omde = -\fr{1}{3}$. Thus, we need to find $\zeta$ for general $\omde$. As we show in both $\omde = -\fr{1}{3}$ and $-1$ cases, $\zeta$ is a function of both $\omde$ and $Q_{\ta}$. Also it depends on $\alpha$. Thus, we are able to derive a reasonable fitting formula for $\zeta$ as a function of $\omde$, $Q_{\ta}$, and $\alpha$. When we investigate the numerical solutions of Eqs. (\ref{ddy}) for general $\omde$, we obtain the fitting formula of $\zeta$ \be \zeta_{\sk} = \Bigl( \fr{3 \pi}{4} \Bigr)^2 \Omega_{\rm{mta}}^{-0.724 + 0.157 \Omega_{\rm{mta}} + \alpha (1 + \omde)(1+3\omde) (0.064 - 0.368 \Omega_{\rm{mta}})} \, , \label{zetaSK} \ee where $\Omega_{\rm{mta}} = \Omega_{\rm{m}}(x_{\ta}) = \fr{1}{1 + Q_{\ta}^{-1}}$ and it becomes $1$ when $\Omo = 1$. Thus, this choice of the fitting formula is consistent with the fact that $\zeta = (\fr{3\pi}{4})^2$ in the EdS universe. Also $\Omega_{\rm{mta}}$ is a function of $\omde$, $z_{\ta}$, and $\Omo$, and thus it is a function of all the variables in the consideration. Even though we consider $\alpha = 1$ model in this paper, the above fitting formula is good for any value of $\alpha$. The special choice of this fitting form of $\zeta$ is independent of $\alpha$ when $\omde = -\fr{1}{3}$ or $-1$ in order to be consistent with real models. In Ref. \cite{WS}, $\zeta$ is given by another form for the fitting formula, \be \zeta_{\ws} = \Bigl( \fr{3 \pi}{4} \Bigr)^2 \Omega_{\rm{mta}}^{-0.79 + 0.26 \Omega_{\rm{mta}} - 0.06 \omde} \, . \label{zetaWS} \ee In Fig. \ref{fig3}, we compare the values of $\zeta_{\sk}$ and $\zeta_{\ws}$ for $\omde = -\fr{1}{3}$ with the exact $\zeta_{13}$ given in Eq.(\ref{zeta}) for the different values of $\Omo$ and $z_{\ta}$. We use $z_{\ta} = 1$ and normalize $\zeta$s with $(\fr{3 \pi}{4})^2$ in the left panel of Fig. \ref{fig3}. The dashed line depicts $\zeta_{\ws}$ and the solid and dot-dashed lines describe $\zeta_{13}$ and $\zeta_{\sk}$, respectively. $\zeta_{13}$ and $\zeta_{\sk}$ are too close to each other and look like a single line. We show $\zeta$s as a dependence on $z_{\ta}$ when we choose $\Omo = 0.3$ in the right panel of Fig. \ref{fig3}. Again both $\zeta_{13}$ and $\zeta_{\sk}$ are very close to each other in most of the $z_{\ta}$ range. We will show later that the fitting formula given in Eq. (\ref{zetaSK}) shows the better fitting than that of Eq (\ref{zetaWS}) for general $\omde$.

As given in Eq. (\ref{ddy2}), $y$ should be a function of $\omde$, $Q_{\ta}$, and $\zeta$. Even though the function of $y$ for the general value of $\omde$ should not be the same as Eq. (\ref{y133}), we modify this equation to obtain the approximate analytic solution of $y$ as \ba && \sqrt{y(1-y)} - \AS [\sqrt{y} ] + \fr{\pi}{2}  + \Biggl( -\fr{(1+3\omde)}{3} \Omega_{\rm{mta}} \Biggr)^{2.7 + 0.1 \Omega_{\rm{mta}} -3.4 (1 + \omde) -0.02 \zeta_{\sk} + 3(z_{\ta} - 0.6)} \nonumber \\ && = \sqrt{\zeta_{\sk}} ( \tau - \tau_{\ta}) \label{yanal} \\ && = \fr{2}{3} \sqrt{\zeta_{\sk}} \Biggl( x^{\fr{3}{2}} F \Bigl[ -\fr{1}{2 \omde}, \fr{1}{2}, 1 - \fr{1}{2 \omde}, - \fr{x^{-3 \omde}}{Q_{\ta}} \Bigr] - F \Bigl[ -\fr{1}{2 \omde}, \fr{1}{2}, 1 - \fr{1}{2 \omde}, - \fr{1}{Q_{\ta}} \Bigr] \Biggr) \, . \nonumber \ea This approximate solution is obtained from interpolating $y$ values for the wide ranges of the cosmological parameters and one can further work on the similar fitting form by changing interested ranges of the cosmological parameters. There are several reasons for the specific choice of the correction term $\Biggl( -\fr{(1+3\omde)}{3} \Omega_{\rm{mta}} \Biggr)^{2.7 + 0.1 \Omega_{\rm{mta}} -3.4 (1 + \omde) -0.02 \zeta_{\sk} + 3(z_{\ta} - 0.6)}$ in the above equation. First, we know that the first two terms in Eq. (\ref{yanal}) are exact when $\omde = -\fr{1}{3}$. Thus, the correction term should not appear for $\omde = -\fr{1}{3}$ and thus we put $(1+3\omde)$ in this correction. Also as $z_{\ta}$ increases, the contribution of the correction term decreases. Thus, in the approximate solution we have the power $3(z_{\ta} - 0.6)$ term. Also we find that $z_{\ta} \sim 0.6$ in order to have $z_{\vir} \sim 0$ in the concordance model.
\begin{center}
\begin{figure}
\vspace{1.5cm}
\centerline{
\psfig{file=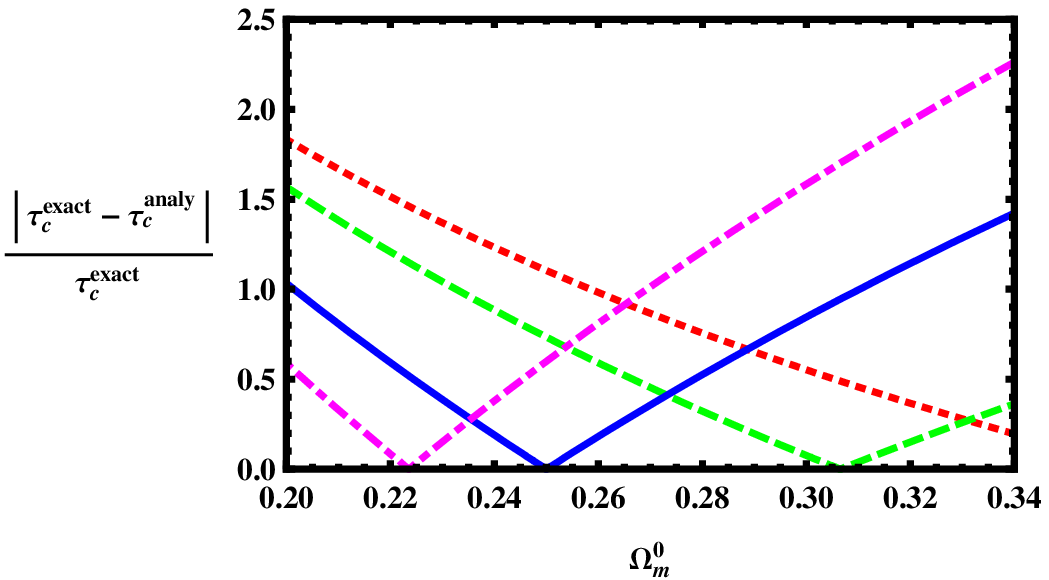, width=6.5cm} \psfig{file=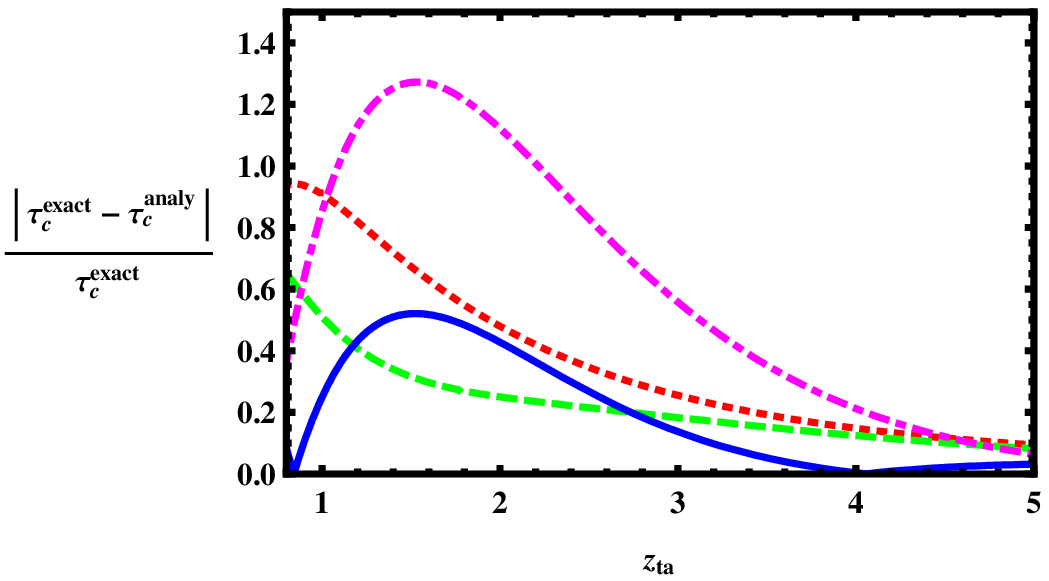, width=6.5cm}}
\vspace{-0.5cm}
\caption{ Error (\%) of the analytic solution of $\tau_{c}$ obtained from Eq. (\ref{yanal}). a) Error of $\tau_{c}^{\analy}$ against $\Omo$ when $z_{\ta} = 1.1 (z_{\vir} \sim 0.3)$ for the different $\omde = -0.8$, $-0.9$, $-1.0$, and $-1.05$ models, denoted by the dotted, dashed, solid, and dot-dashed lines, respectively. b) Error of $\tau_{c}^{\analy}$ against $z_{\ta}$ when $\Omo = 0.27$ for the different $\omde$ models.} \label{fig4}
\end{figure}
\end{center}

Since we know the exact value of $\tau_{c}$ from Eq. (\ref{xtau}), we are able to estimate the goodness of the approximate solution given in Eq. (\ref{yanal}). The errors of $\tau_{c}$ for the different models obtained from Eq. (\ref{yanal}) are shown in Fig. \ref{fig4}. We show the errors of $\tau_{c}^{\analy}$ against $\Omo$ with $z_{\ta} = 1.1$ for the different $\omde$ models in the left panel of Fig. \ref{fig4}. The dotted, dashed, solid, and dot-dashed lines correspond to $\omde = -0.8$, $-0.9$, $-1.0$, and $-1.05$, respectively. If we limit $0.24 \leq \Omo \leq 0.3$, then the errors are less than $1.6$ \% for all the models. We also depict the errors of $\tau_{c}^{\analy}$ against $z_{\ta}$ with $\Omo = 0.27$ for the different $\omde$ models in the right panel of Fig. \ref{fig4}. Again, the dotted, dashed, solid, and dot-dashed lines correspond to $\omde = -0.8$, $-0.9$, $-1.0$, and $-1.05$, respectively. The errors of $\tau_{c}^{\analy}$ for $\omde \geq -1.0$ are less than $1$ \% for the entire range of $z_{\ta}$.

\begin{center}
\begin{figure}
\vspace{1.5cm}
\centerline{
\psfig{file=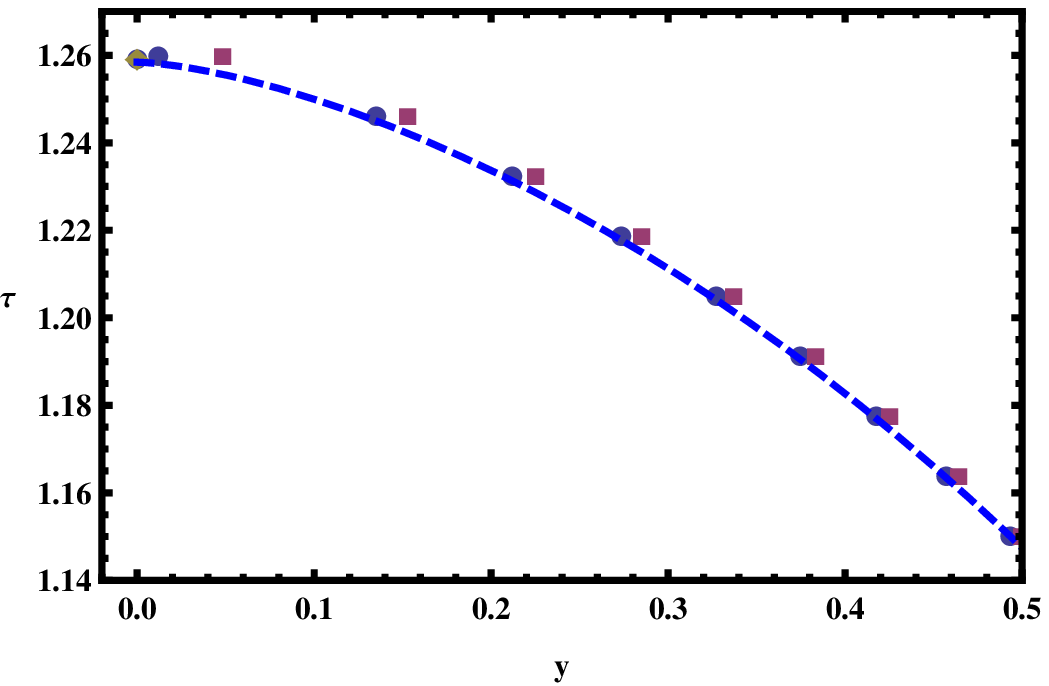, width=6.5cm} \psfig{file=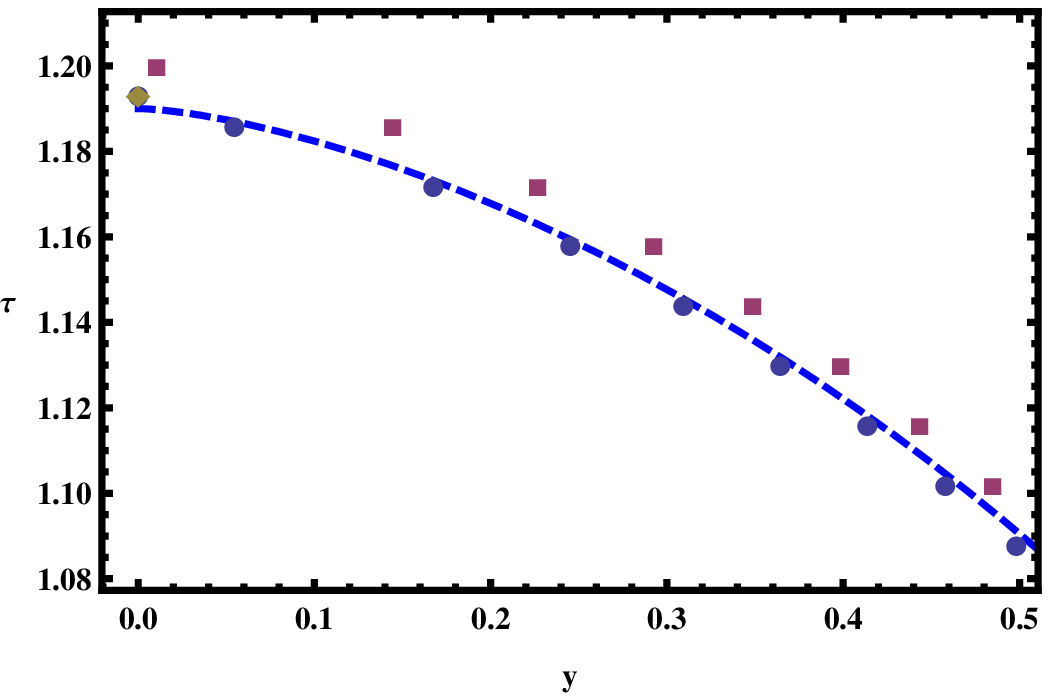, width=6.5cm} }
\vspace{-0.5cm}
\caption{ Comparing numerical solutions of $y$ when we use $\zeta_{\ws}$ and $\zeta_{\sk}$ with $y_{\analy}$ for the different models. a) $\tau$ v.s. $y$ when $\omde = -0.9$, $\Omo = 0.3$ and $z_{\ta} = 1$. b) $\tau$ v.s. $y$ when $\omde = -1.0$, $\Omo = 0.3$, and $z_{\ta} = 0.4$. The circle and rectangular points represent the numerical $y$ values when we use $\zeta_{\sk}$ and $\zeta_{ws}$, respectively. The diamond points at $y = 0$ indicate the exact values of $\tau_{c}$. } \label{fig5}
\end{figure}
\end{center}

We claim that $\zeta_{\sk}$ gives the better fit than $\zeta_{\ws}$ for both general $\omde$ and a wide range of the cosmological parameters. As we expect, both give a good approximate value when $z_{\ta}$ increases. Again this comparison is possible because we know the exact collapsing time $\tau_{c} = 2 \tau_{\ta}$ for each $\omde$ model. In Fig. \ref{fig5}, we demonstrate the evolution of both numerical and approximate analytic solutions $y$ for some cases. The circle and rectangular points represent the numerical $y$ values when we use $\zeta_{\sk}$ and $\zeta_{\ws}$, respectively. The diamond point (the grey point at $y = 0$) indicates the exact value of $\tau_{c}$ obtained from Eq. (\ref{xtau}). The dashed line depicts the approximate solution of $y$ given in Eq. (\ref{yanal}). In the left panel of Fig. \ref{fig5}, we show both numerical and analytic solutions when $\omde = -0.9$, $\Omo = 0.3$, and $z_{\ta} = 1.0$. We also demonstrate them when $\omde = -1.0$, $\Omo = 0.3$, and $z_{\ta} = 0.4$ in the right panel of Fig. \ref{fig5}. Even though we show $z_{\ta} = 0.4$ case in this figure, we will show in the next section that there exists the minimum $z_{\ta}$ for each model when we use the virial theorem.
\section{Virialization}
\setcounter{equation}{0}
If only the matter virializes, then we are not able to use the energy conservation in the virial theorem because the dark energy does not virialize with the dark matter \cite{0505308,0507195,0608030,0610184}. \be U_{\mc}(z_{\ta}) + U_{\de}(z_{\ta}) \, \geq \, \fr{1}{2} U_{\mc}(z_{\vir}) + 2 U_{\de}(z_{\vir}) \, , \label{vir} \ee where $U_{\mc}$ and $U_{\de}$ are the potential energies associated with the spherical mass overdensity and with the DE, respectively. Thus, in the presence of the smooth DE, the dark matter can reach a quasi-equilibrium state where virialization holds instantaneously.  To estimate the virial radius $y_{\vir} = \fr{R_{\vir}}{R_{\ta}}$ we use the assumption $\rho_{\de}(z_{\vir}) = \rho_{\de}(z_{\ta})$ as in the reference \cite{0507195}. With this approximation, we are able to obtain the equation of $y_{\vir} = \fr{R_{\vir}}{R_{\ta}}$ for any value of $\omde$ : \be 4 (1 + 3 \omde) y_{\vir}^3 - 2 \Bigl[ \zeta Q_{\ta} + (1 + 3 \omde) \Bigr] y_{\vir} + \zeta Q_{\ta} = 0 \, . \label{yvir} \ee An interesting feature in this equation is that we obtain $y_{\vir} = \fr{1}{2}$ like that in the EdS universe when $\omde = -\fr{1}{3}$ for any given values of $\Omo$ and $z_{\ta}$ \cite{sky}. Thus, we are able to obtain $x_{\vir}$ of the $\omde = -\fr{1}{3}$ model for the different values of $\Omo$ and $z_{\ta}$ from Eq. (\ref{y133}). With simple algebra by using Eqs. (\ref{xtau}), (\ref{tautaw}), (\ref{y133}), and (\ref{zeta}), we obtain \ba && x_{\vir}^{\fr{3}{2}} F \Bigl[\fr{1}{2}, \fr{3}{2}, \fr{5}{2}, -\fr{x_{\vir}}{Q_{\ta 13}} \Bigr] = \Bigl( \fr{1}{\pi} + \fr{3}{2} \Bigr) F \Bigl[ \fr{1}{2}, \fr{3}{2}, \fr{5}{2}, -\fr{1}{Q_{\ta 13}} \Bigr] \nonumber \\  && = \fr{3}{2} Q_{\ta 13}^{\fr{3}{2}} \Biggl( \sqrt{ \fr{x_{\vir}}{Q_{\ta 13}} \Bigl(1 + \fr{x_{\vir}}{Q_{\ta 13}} \Bigr)} - \ln \Bigl[ \sqrt{\fr{x_{\vir}}{Q_{\ta 13}}} + \sqrt{1 + \fr{x_{\vir}}{Q_{\ta 13}}}\Bigr] \Biggr) \, , \label{xvirw13} \ea where we use the relations $\fr{x_{\vir}}{Q_{\ta}}  F \Bigl[ \fr{1}{2}, \fr{3}{2}, \fr{5}{2}, -\fr{x_{\vir}}{Q_{\ta}} \Bigr] = \fr{3}{2} \Bigl( 1 + \fr{x_{\vir}}{Q_{\ta}} \Bigr)  F \Bigl[ \fr{1}{2}, \fr{3}{2}, \fr{3}{2}, -\fr{x_{\vir}}{Q_{\ta}} \Bigr] - \fr{3}{2}  F \Bigl[\fr{1}{2}, \fr{1}{2}, \fr{3}{2}, -\fr{x_{\vir}}{Q_{\ta}} \Bigr]$, $F \Bigl[\fr{1}{2}, \fr{3}{2}, \fr{3}{2}, -\fr{x_{\vir}}{Q_{\ta}} \Bigr] = \Bigl( 1 + \fr{x_{\vir}}{Q_{\ta}}\Bigr)^{-\fr{1}{2}}$, and $F \Bigl[\fr{1}{2}, \fr{1}{2}, \fr{3}{2}, -\fr{x_{\vir}}{Q_{\ta}} \Bigr] = \Bigl( \sqrt{\fr{x_{\vir}}{Q_{\ta}}} \Bigr)^{-1} \ln \Bigl[ \sqrt{\fr{x_{\vir}}{Q_{\ta}}} + \sqrt{1 + \fr{x_{\vir}}{Q_{\ta}}} \Bigr]$ in the second equality \cite{Abramowitz}. Even though we obtain the analytic expression of $x_{\vir}$ in Eq. (\ref{xvirw13}), generally this equation can be solved in a non-algebraic way. However, $x_{\vir}$ can be represented by the approximate analytic form for the wide ranges of the cosmological parameters ($\Omo$ and $z_{\ta}$) if one analyzes the form by interpolating $x_{\vir}$ values. 

For the EdS universe, Eqs. (\ref{xtau}), (\ref{y133}), and (\ref{zeta}) become \ba && \fr{2}{3} x^{\fr{3}{2}} = \tau \, , \label{xtauEdS} \\ && \sqrt{y(1-y)} - \AS [\sqrt{y} ] + \fr{\pi}{2} = \sqrt{\zeta} ( \tau - \fr{2}{3} ) = \fr{\pi}{2} \Bigl(x^{\fr{3}{2}} - 1 \Bigr) \, ,  \label{y133EdS} \\ && \zeta = (\fr{3 \pi}{4})^2 \label{zetaEdS} \, ,\ea
where we use $F[\fr{1}{2},-\fr{1}{2\omde},1-\fr{1}{2\omde},0] = 1$ and $\tau_{\ta} = \fr{2}{3}$. The above equations can be generalized to any value of $\omde$ when $\Omo$ approaches to $1$ or $z_{\ta}$ goes to a very high value (for most of the parameter spaces $z_{\ta} \geq 5$). In the EdS universe, $y_{\vir} = \fr{1}{2}$ from Eq. (\ref{yvir}). The commonly used assumption to obtain the nonlinear overdensity $\Delta_{c}$ in the EdS universe is that $\tau_{c} = 2 \tau_{\ta} = \fr{4}{3}$ which is the collapsing time for $y_{c} =0$ even though one uses $y_{\vir} = \fr{1}{2}$ to obtain $\Delta_{c}$. By using this assumption, one obtain $x_{c} = 2^{\fr{2}{3}}$ from Eq. (\ref{xtauEdS}). Thus, one obtains \be \Delta_{c}^{\EdS} = \zeta \Biggl( \fr{x_{c}}{y_{\vir}} \Biggr)^3 = \Biggl( \fr{3 \pi}{4} \Biggr)^2 \Biggl( \fr{2^{\fr{2}{3}}}{2^{-1}} \Biggr)^{3} = 18 \pi^2 \simeq 178 \, . \label{DeltacEdS1} \ee
However, we can obtain the correct value of $\Delta_{\vir}$ for the EdS universe because we know the exact relations between $\tau$, $x$, and $y$.  After replacing $y_{\vir} = \fr{1}{2}$ into Eq. (\ref{y133EdS}), we obtain the virialization time $\tau_{\vir} = \fr{2}{3 \pi} + 1 < \tau_{c}$. From this correct value of $\tau_{\vir}$, we obtain $x_{\vir} = (\fr{1}{\pi} + \fr{3}{2})^{\fr{2}{3}}$ by using Eq. (\ref{xtauEdS}). Thus, the correct value of the non-linear overdensity $\Delta_{\vir}$ for the EdS universe is \cite{sky} \be \Delta_{\vir}^{\EdS} = \zeta \Biggl( \fr{x_{\vir}}{y_{\vir}} \Biggr)^3 = \Biggl( \fr{3 \pi}{4} \Biggr)^2 \Biggl( \fr{\Bigl(\fr{1}{\pi} + \fr{3}{2} \Bigr)^{\fr{2}{3}}}{2^{-1}} \Biggr)^{3} = 18 \pi^2 \Biggl(\fr{1}{2 \pi} + \fr{3}{4} \Biggr)^2 \simeq 147 \, . \label{DeltavirEdScor} \ee As $\omde$ becomes large, the structure is formed earlier and $\Delta_{\vir}$ becomes large. There is only gravity involved in the structure formation in the EdS universe and thus it requires less overdensity than the one with the pressure. Thus, the minimum overdensity for general $\omde$ models is about $147$ instead of $178$. We show this in Tables \ref{table1} and \ref{table2}. As $z_{\ta}$ increases, $\Delta_{\vir}$ approaches to $147$ instead of $178$.

The ratio of cluster to background density at the virialized epoch $z_{\vir}$ for $\omde = -\fr{1}{3}$ becomes \be \Delta_{\vir} \equiv \Delta(z_{\vir}) = \fr{\rho_{\mc}}{\rho_{\rm{m}}} \Biggl|_{z_{\vir}} = \zeta \Biggl( \fr{x_{\vir}}{y_{\vir}} \Biggr)^3 = 18 \pi^2 \fr{x_{\vir}^3}{4 (F \Bigl[\fr{1}{2}, \fr{3}{2}, \fr{5}{2}, -\fr{x_{\vir}}{Q_{\ta}} \Bigr])^2} \, , \label{Deltacw13} \ee where we use Eqs. (\ref{rhomdot}), (\ref{rhocdot}), and (\ref{zeta}). We also use the fact that $y_{\vir} = \fr{1}{2}$ independent of $Q_{\ta}$. We are able to obtain $\Delta_{\vir}$ for the given cosmological parameters by using Eqs. (\ref{xvirw13}) and (\ref{Deltacw13}). It is also useful to express $z_{\vir}$ by using $z_{\ta}$ and $x_{\vir}$ from Eq. (\ref{x}) \be z_{\vir} = \fr{1 + z_{\ta}}{x_{\vir}} - 1 \, . \label{zvir} \ee  

Now we are able to extend the previous consideration to general values of $\omde$ by using Eqs. (\ref{xtau}), (\ref{zetaSK}), (\ref{yanal}), and (\ref{yvir}). We show several quantities for the different models in Tables \ref{table1} and \ref{table2}. We need to clarify several things in these tables. First, structures start to form earlier for lager $\omde$ and thus $\Delta_{\vir}$ rises with increasing $\omde$. 
Thus, we are able to distinguish between the models from the observed cluster at a given $z_{\vir}$. The other important feature is that as $\Omo$ and $z_{\ta}$ increase, the model dependence of the observed quantity $\Delta_{\vir}$ decreases. Thus, the low-redshift clusters are the better objects to be used for studying the properties of DE. Again as $z_{\ta}$ increases, $\Delta_{\vir}$ approaches to $147$ instead of $18 \pi^2 \simeq 178$.

\begin{center}
    \begin{table}
    \begin{tabular}{ | c | c | c | c | c | c | c | c | c | c | c | c | c | c | c | c | c |}
    \hline
      &  \multicolumn{8}{|c|}{$z_{\ta} = 0.7$} & \multicolumn{8}{|c|}{$z_{\ta} = 1.1$} \\ \cline{2-17}
     $\omde$ & \multicolumn{4}{|c|}{$\Omo = 0.24$($z_{\vir} \sim 0.02$) } & \multicolumn{4}{|c|}{$\Omo = 0.3$($z_{\vir} \sim 0.04$) } &  \multicolumn{4}{|c|}{$\Omo = 0.24$($z_{\vir} \sim 0.31$)} & \multicolumn{4}{|c|}{$\Omo = 0.3$($z_{\vir} \sim 0.33$)} \\ \cline{2-17}
    &$\zeta_{\sk}$ & $x_{\vir}$  & $y_{\vir}$ & $\Delta_{\vir}$ & $\zeta_{\sk}$ & $x_{\vir}$ & $y_{\vir}$ & $\Delta_{\vir}$ & $\zeta_{\sk}$ & $x_{\vir}$ & $y_{\vir}$ & $\Delta_{\vir}$ & $\zeta_{\sk}$ & $x_{\vir}$ & $y_{\vir}$ & $\Delta_{\vir}$  \\ \hline
    $-\fr{1}{3}$ & $11.2$ & $1.62$ & $0.50$ & $384$ & $9.8$ & $1.60$ & $0.50$ & $322$
           & $10.2$ & $1.61$ & $0.50$ & $339$ & $9.0$ & $1.59$ & $0.50$ & $289$ \\ \hline
    $-0.6$ & $9.1$ & $1.65$ & $0.53$ & $282$ & $8.2$ & $1.63$ & $0.52$ & $250$
           & $8.0$ & $1.61$ & $0.52$ & $236$ & $7.3$ & $1.59$ & $0.52$ & $214$ \\ \hline
    $-0.8$ & $8.1$ & $1.66$ & $0.53$ & $244$ & $7.4$ & $1.64$ & $0.53$ & $222$
           & $7.1$ & $1.60$ & $0.52$ & $201$ & $6.7$ & $1.58$ & $0.52$ & $188$ \\ \hline
    $-0.9$ & $7.8$ & $1.66$ & $0.54$ & $231$ & $7.2$ & $1.64$ & $0.53$ & $213$
           & $6.8$ & $1.60$ & $0.53$ & $193$ & $6.5$ & $1.58$ & $0.52$ & $182$ \\ \hline
    $-1.0$ & $7.6$ & $1.66$ & $0.54$ & $222$ & $7.1$ & $1.65$ & $0.53$ & $212$
           & $6.6$ & $1.60$ & $0.53$ & $189$ & $6.4$ & $1.59$ & $0.52$ & $181$ \\ \hline
    \end{tabular}
    \caption{$\zeta_{\sk}$, $x_{\vir}$, $y_{\vir}$, and $\Delta_{\vir}$ with the given values of the cosmological parameters for the different DE ($\omde$) models. As $\omde$ becomes large, the structure is formed earlier and $\Delta_{\vir}$ becomes large. }
    \label{table1}
    \end{table}
\end{center}

\begin{center}
    \begin{table}
    \begin{tabular}{ | c | c | c | c | c | c | c | c | c | c | c | c | c | c | c | c | c |}
    \hline
      &  \multicolumn{8}{|c|}{$z_{\ta} = 1.4$} & \multicolumn{8}{|c|}{$z_{\ta} = 5.0$} \\ \cline{2-17}
     $\omde$ & \multicolumn{4}{|c|}{$\Omo = 0.24$($z_{\vir} \sim 0.53$)} & \multicolumn{4}{|c|}{$\Omo = 0.3$($z_{\vir} \sim 0.54$)} &  \multicolumn{4}{|c|}{$\Omo = 0.24$($z_{\vir} \sim 3.0$)} & \multicolumn{4}{|c|}{$\Omo = 0.3$($z_{\vir} \sim 3.0$)} \\ \cline{2-17}
    &$\zeta_{\sk}$ & $x_{\vir}$  & $y_{\vir}$ & $\Delta_{\vir}$ & $\zeta_{\sk}$ & $x_{\vir}$ & $y_{\vir}$ & $\Delta_{\vir}$ & $\zeta_{\sk}$ & $x_{\vir}$ & $y_{\vir}$ & $\Delta_{\vir}$ & $\zeta_{\sk}$ & $x_{\vir}$ & $y_{\vir}$ & $\Delta_{\vir}$  \\ \hline
    $-\fr{1}{3}$ & $9.6$ & $1.60$ & $0.50$ & $316$ & $8.6$ & $1.58$ & $0.50$ & $271$
           & $7.2$ & $1.55$ & $0.50$ & $215$ & $6.8$ & $1.50$ & $0.50$ & $183$ \\ \hline
    $-0.6$ & $7.4$ & $1.59$ & $0.52$ & $216$ & $6.9$ & $1.57$ & $0.51$ & $198$
           & $5.9$ & $1.52$ & $0.50$ & $161$ & $5.8$ & $1.50$ & $0.50$ & $154$ \\ \hline
    $-0.8$ & $6.7$ & $1.57$ & $0.52$ & $184$ & $6.4$ & $1.56$ & $0.52$ & $176$
           & $5.7$ & $1.50$ & $0.50$ & $151$ & $5.6$ & $1.50$ & $0.50$ & $150$ \\ \hline
    $-0.9$ & $6.4$ & $1.57$ & $0.52$ & $178$ & $6.2$ & $1.55$ & $0.51$ & $171$
           & $5.6$ & $1.50$ & $0.50$ & $150$ & $5.6$ & $1.49$ & $0.50$ & $147$ \\ \hline
    $-1.0$ & $6.3$ & $1.57$ & $0.52$ & $175$ & $6.1$ & $1.56$ & $0.51$ & $170$
           & $5.6$ & $1.50$ & $0.50$ & $150$ & $5.6$ & $1.49$ & $0.50$ & $147$ \\ \hline
    \end{tabular}
    \caption{$\zeta_{\sk}$, $x_{\vir}$, $y_{\vir}$, and $\Delta_{\vir}$ with the given values of the cosmological parameters for the different DE ($\omde$) models. Independent of model parameters, $\Delta_{\vir}$ converges to 147 as $z_{\ta}$ increases.}
    \label{table2}
    \end{table}
\end{center}

From Eq. (\ref{Deltacw13}), we are able to investigate the linear perturbation at early epoch $\Delta \,\, \stackrel{\tau\rightarrow 0}{\longrightarrow} \,\, 1 + \delta_{\rm{lin}}$ \cite{sky} \be \Delta \equiv 1 + \delta_{\rm{NL}} \stackrel{\tau\rightarrow 0}{\longrightarrow} \, 1 + \delta_{\rm{lin}} = 1 + \fr{3}{5} \Bigl(\sqrt{\zeta} \fr{3}{2} \tau \Bigr)^{\fr{2}{3}} = 1 + \fr{3}{5} \Bigl(\sqrt{\zeta} \sqrt{\Omega_{\rm{mta}}} \fr{t}{t_{\ta}} \Bigr)^{\fr{2}{3}} \, , \label{Deltadelta} \ee where we use $H_{\ta} = \fr{2}{3} t_{\ta}^{-1}$ in the early (matter-dominated) epoch. This is equal to the famous result for the EdS universe ($\sqrt{\zeta} = \fr{3 \pi}{4}$ and $\Omega_{\rm{mta}} = 1$), $\delta_{\rm{lin}} = \fr{3}{5} \Bigr( \fr{3 \pi}{4} \fr{t}{t_{\ta}} \Bigr)^{\fr{2}{3}}$ \cite{Peebles,sky}. As $\tau$ approaches to zero, the exact solutions of $x$ and $y$ given in Eqs. (\ref{xtau}) and (\ref{y133l}) become the good approximate solutions for general values of $\omde$. In this limit, $y_{\vir} \simeq \fr{1}{2}$ and thus $\tau_{\vir} \simeq \Bigl(\fr{3}{2} + \fr{1}{\pi} \Bigr) \tau_{\ta}$ from Eq.~(\ref{y133l}). Thus, \be \delta_{\rm{lin}}(z_{\vir}) = \fr{3}{5} ( \sqrt{\zeta})^{\fr{2}{3}} \Biggl( \Bigl( \fr{3}{4} + \fr{9 \pi}{8} \Bigr) \fr{1}{\sqrt{\zeta}} \Biggr)^{\fr{2}{3}} = \fr{3}{20} (6 + 9 \pi)^{\fr{2}{3}} \simeq 1.58 \, , \label{deltavir} \ee where we use $\sqrt{\zeta} = \fr{\pi}{2 \tau_{\ta}}$. The above result given in Eq. (\ref{deltavir}) is true for any value of $\omde$ because $\omde$ dependence of $x$ and $y$ is disappeared in the $\tau \rightarrow 0$ limit. From this critical density threshold $\delta_{\rm{lin}}(z_{\vir})$, we are able to obtain $\delta_{\rm{lin}}$ at any epoch by using the relation \be \delta_{\rm{lin}}(z) = \fr{D_{g}(z)}{D_{g}(z_{\vir})} \delta_{\rm{lin}}(z_{\vir}) \, , \label{deltaz} \ee where $D_{g}$ is the linear growth factor. There is the exact analytic form of $D_{g}$ for the DE model with the constant eos \cite{SK}.

From the analytic forms of dynamical quantities $x$, $y$, and $\zeta$, we are able to estimate the abundances of both virialized and non-virialized clusters at any epoch. Also the temperature and luminosity functions are able to be computed at any epoch \cite{SN}. Thus, these analytic forms provide a very accurate and useful tool for probing the properties of DE.

\section{Conclusions}
\setcounter{equation}{0}

The exact solution for the evolution of the background scale factor in a flat universe with the dark energy is given as a function of time. This provides the exact value of collapsing time and makes it possible to test the accuracies of both the analytic and the numerical solutions. We obtain the exact and approximate solutions of the evolution of the overdensity radius with high accuracy. From these solutions we are able to estimate the non-linear overdensity at any epoch.

Even though the analytic forms of $y$ and $\zeta$ are obtained for the constant $\omde$ models, they can be generalized to the slowly varying $\omde \geq -\fr{1}{3}$ because they provide the very accurate values of them for the wide range of cosmological parameters.

The virial theorem provides the exact value of the virial epoch for given models. From this, the value of the non-linear overdensity is obtained. In the high matter density and/or the high-redshift clusters, the model dependence of the observed quantities becomes weaker. We are able to categorize the physical quantities of the low and high redshift clusters for the given models.

From the correct value of the virial radius of an EdS universe, we obtain the correct linear and non-linear overdensities for an EdS universe, which are lower than the known values. In this paper, we show that clusters are formed earlier than those from the conventional model. Thus, the mass of the cluster is larger than one obtained from the conventional model. This will affect to the prediction for the different models for the cluster abundances, the weak gravitational lensing, etc.

\appendix
\section{Appendix}
\setcounter{equation}{0}
We derive the exact solution of $\tau$ as a function of $x$ given in Eq. (\ref{xtau}). After replacing the variables $Z= - \fr{x^{-3 \omde}}{Q_{\ta}}$ and $T = (\fr{x'}{x})^{-3\omde}$, Eq. (\ref{xtau}) becomes \cite{Abramowitz} \be \int_{0}^{x} \fr{dx'}{\sqrt{x'^{-1} + \fr{x'^{-3\omde -1}}{Q_{\ta}}}} = - \fr{x^{\fr{3}{2}}}{3 \omde} \int_{0}^{1} T^{-1-\fr{1}{2\omde}} (1 - ZT)^{-\fr{1}{2}} dT = \fr{2}{3} x^{\fr{3}{2}} F \Bigl[ \fr{1}{2}, -\fr{1}{2\omde}, 1- \fr{1}{2 \omde}, -\fr{x^{-3\omde}}{Q_{\ta}} \Bigr] \, , \label{xtaut} \ee where $F$ is the hypergeometric function and we use the gamma function relation $\Gamma[1+b] = b \Gamma[b]$.

The evolution equation of $\tau$ as a function of $y$ when $\omde = -\fr{1}{3}$ is given in Eq. (\ref{y13}). The exact analytic solution of this equation is known \cite{sky} : \ba && \AS [\sqrt{y} ] - \sqrt{y(1-y)} = \sqrt{\zeta_{13}} \tau \,\, , \rm{when} \,\, \tau \leq \tau_{\ta 13} \, , \label{y133appl} \\&& \sqrt{y(1-y)} - \AS [\sqrt{y} ] + \fr{\pi}{2} = \sqrt{\zeta_{13}} ( \tau - \tau_{\ta 13} ) \,\, , \rm{when} \,\, \tau \geq \tau_{\ta 13} \, , \label{y133appg} \ea 
where $\zeta_{13}$ is given in Eq. (\ref{zeta}) and $\tau_{\ta 13}$ is $\tau_{\ta}$ for $\omde = -\fr{1}{3}$. $\tau_{\ta 13}$ is obtained from Eq. (\ref{tautaw}) as \be \tau_{\ta 13} = \fr{2}{3} F \Bigl[ \fr{1}{2}, \fr{3}{2}, \fr{5}{2}, - \fr{1}{Q_{\ta 13}} \Bigr] \, , \label{tautaw13} \ee where $Q_{\ta 13}$ is $Q_{\ta}$ for  $\omde = -\fr{1}{3}$. 

The equation governing $y$ for $\omde = -1$ is given by Eq. (\ref{ddywL}) with $c_{1} = \zeta + Q_{\ta}^{-1}$. \be \int^{y} \fr{dy'}{\sqrt{ \zeta (\fr{1}{y} -1) + \fr{1}{Q_{\ta}} (y^2 -1)}} = c_{2} \pm \tau \, . \label{dywLapp} \ee The above equation (\ref{dywLapp}) is solved analytically by \cite{Abramowitz} \ba && \int^{y} \fr{dy'}{\sqrt{ \zeta (\fr{1-y}{y}) + Q_{\ta}^{-1}(y^2-1)}} = - \fr{2 \sqrt{2 Q_{\ta}}}{\sqrt{2 Q_{\ta} \zeta - 1 + \sqrt{1 + 4 Q_{\ta} \zeta}}} \Biggl( \EF \Bigl[\AS [B], C \Bigr] \nonumber \\ && - \fr{2 (Q_{\ta} \zeta - 2)}{(3 + \sqrt{1 + 4 Q_{\ta} \zeta}) } \EPi \Bigl[ A, \AS [B], C \Bigr] \Biggr) \,\,\, \rm{where} \,\, A = \fr{2 \sqrt{1 + 4Q_{\ta} \zeta}}{3 + \sqrt{1 + 4 Q_{\ta} \zeta}} \, , \nonumber \\ &&  B = \fr{\sqrt{3 + \fr{(2 Q_{\ta} \zeta -4)}{(1-y)}  + \sqrt{1 + 4 Q_{\ta} \zeta}}}{\sqrt{2(1 + 4 Q_{\ta} \zeta)}} \, , C = \fr{2 \sqrt{1 + 4 Q_{\ta} \zeta}}{-1 + 2 Q_{\ta} \zeta + \sqrt{1+4Q_{\ta} \zeta}} \, , \label{ywLapp} \ea where EllipticF and EllipticPi represent the elliptic integral of the first kind and the incompelete elliptic integral of the third kind, respectively.

We also check the linear perturbation at early epoch, $\Delta \,\, \stackrel{\tau\rightarrow 0}{\longrightarrow} \,\, 1 + \delta_{\rm{lin}}$. We use Eq. (\ref{y133appl}) because we consider $\tau \rightarrow 0$. Even though this solution is exact only for $\omde = -\fr{1}{3}$, it is safe for us to use it for the general values of $\omde$ because the second term in Eq. (\ref{ddy2}) becomes zero in $y \rightarrow 0$ limit. Thus, $\delta_{\rm{lin}}$ obtained from this limit is independent of the value of $\omde$. $\fr{2}{3} x^{\fr{3}{2}} \simeq \tau$ and $\fr{2}{3}y^{\fr{3}{2}} + \fr{1}{5} y^{\fr{5}{2}} \simeq \sqrt{\zeta} \tau$ as $\tau \rightarrow 0$. Thus we obtain \be \Delta \equiv 1 + \delta_{\rm{NL}} = \zeta \Bigl( \fr{x}{y} \Bigr)^3 \, \stackrel{\tau\rightarrow 0}{\longrightarrow} \, \zeta \Biggl( \fr{1}{\sqrt{\zeta}} \Bigl( 1 + \fr{3}{10} y \Bigr) \Biggr)^2 \simeq 1 + \fr{3}{5} y \,\, \Rightarrow \,\, \delta_{\rm{NL}} \rightarrow \delta_{\rm{lin}} = \fr{3}{5} \zeta^{\fr{1}{3}} \Bigl( \fr{3}{2} \tau \Bigr)^{\fr{2}{3}} \, . \label{Deltadeltaapp} \ee

We thank S.~Basilakos and D.~F.~Mota for useful comments.

\end{document}